\documentclass[twocolumn,showpacs,preprintnumbers,amsmath,amssymb]{revtex4}

\usepackage{graphicx}
\usepackage{dcolumn}
\usepackage{bm}

\begin{document}

\title{Spatiotemporal dynamics of quantum jumps with Rydberg atoms}

\author{Tony E. Lee}
\author{M. C. Cross}
\affiliation{Department of Physics, California Institute of Technology, Pasadena, California 91125, USA}

\date{\today}

\begin{abstract}
We study the nonequilibrium dynamics of quantum jumps in a one-dimensional chain of atoms. Each atom is driven on a strong transition to a short-lived state and on a weak transition to a metastable state. We choose the metastable state to be a Rydberg state so that when an atom jumps to the Rydberg state, it inhibits or enhances jumps in the neighboring atoms. This leads to rich spatiotemporal dynamics that are visible in the fluorescence of the strong transition.
\end{abstract}

\maketitle

\section{Introduction}
Rydberg atoms, which are atoms excited to a high principal quantum number $n$, have long drawn interest because of their exaggerated atomic properties. In recent years, people have been particularly interested in the dipole-dipole interaction between Rydberg atoms, which scales as $n^{11}$ and hence can be very strong. This interaction allows one to study many-body effects in a variety of contexts, such as quantum information processing \cite{lukin01,wilk10,isenhower10,saffman10}, quantum phase transitions \cite{weimer10,pohl10,lesanovsky11}, thermalization of closed quantum systems \cite{olmos09,lesanovsky10}, and nonlinear optics \cite{pritchard10,honer11,gorshkov11,sevincli11}.

Recent works have shown that the Rydberg interaction greatly affects how a group of atoms fluoresce \cite{lee11,lee12,ates12}. When the atoms are laser-excited from the ground state to a Rydberg state and spontaneously decay back to the ground state, there are strong temporal correlations between photon emissions of different atoms. In this paper, we study what happens when the atoms are laser-excited to a low-lying excited state as well as a Rydberg state. This three-level scheme leads to qualitatively different behavior: the atoms develop strong spatial correlations that change on a long time scale.

Our idea is based on quantum jumps of a three-level atom \cite{cook85,cohen86,porrati89,plenio98}. It is well known that an atom driven strongly to a short-lived state and weakly to a metastable state occasionally jumps to and from the metastable state. The jumps are visible in the fluorescence of the strong transition, which exhibits distinct bright and dark periods \cite{nagourney86,sauter86,bergquist86}. 

Here, we consider a one-dimensional chain of many three-level atoms, and we let the metastable state be a Rydberg state, so that a jump of one atom affects its neighbors' jumps via the Rydberg interaction. This leads to rich spatiotemporal dynamics, which are observable by imaging the fluorescence of the strong transition. We observe three types of behaviors, corresponding to different parameter regimes: (i) dark regions are localized but expand and contract on a long time scale; (ii) dark regions diffuse across the system and repel each other; (iii) multiple atoms turn dark and bright in unison.

Previous works studied correlated quantum jumps of atoms in the context of the Dicke model \cite{lewenstein87,skornia01}. They concluded that cooperative effects are very difficult to see experimentally, because the interatomic distance must be much smaller than a wavelength. In contrast, the strong Rydberg interaction here allows the interatomic distance to be much longer than a wavelength. Thus, the atoms develop strong correlations while being individually resolvable.

Section \ref{sec:single} reviews quantum jumps in a single atom. Section \ref{sec:many} introduces the many-body model, and the results are discussed in Secs.~\ref{sec:case1} and \ref{sec:case2}. We give example experimental numbers in Sec.~\ref{sec:experiment}. Details of analytical calculations are provided in Appendices A, B, and C.

\begin{figure}
\centering
\includegraphics[width=3.3 in, trim=50 250 0 -60,clip]{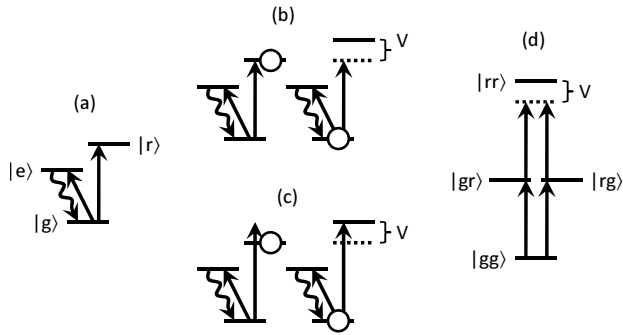}
\caption{\label{fig:levels}(a) An atom has a ground state $|g\rangle$, short-lived excited state $|e\rangle$, and metastable state $|r\rangle$, which is chosen to be a Rydberg state. One observes the spontaneous emission from $|e\rangle$. (b) The $|g\rangle\rightarrow|r\rangle$ transition is originally on resonance ($\Delta_r=0$), but when one atom is in $|r\rangle$, the other atom is off resonance. (c) The $|g\rangle\rightarrow|r\rangle$ transition is originally off resonance ($\Delta_r=V$), but when one atom is in $|r\rangle$, the other atom is on resonance. (d) When $\Delta_r=0$, $|rr\rangle$ is weakly coupled to the other states. Note that (b) and (d) are equivalent.}
\end{figure}

\section{Results for a single atom}\label{sec:single}

We first review quantum jumps in a single atom \cite{cohen86,porrati89,plenio98}. Consider an atom with three levels: ground state $|g\rangle$, short-lived excited state $|e\rangle$, and metastable state $|r\rangle$ [Fig.~\ref{fig:levels}(a)]. In this paper, we choose the metastable state to be a Rydberg state since Rydberg states have long lifetimes \cite{gallagher94}. A laser drives the strong transition ${|g\rangle\rightarrow|e\rangle}$, while another drives the weak transition ${|g\rangle\rightarrow|r\rangle}$. Alternatively, one could use a cascade configuration with ${|e\rangle\rightarrow|r\rangle}$ as the weak transition (see Sec.~\ref{sec:experiment}).

The strong transition acts as a measurement of whether or not the atom is in $|r\rangle$. When the atom is not in $|r\rangle$, the atom is repeatedly excited to $|e\rangle$ and spontaneously emits photons. Occasionally the atom is excited to $|r\rangle$ and stays there, and the fluorescence from the strong transition turns off. Eventually, the atom returns to $|g\rangle$, and the fluoresence turns back on. Thus, the fluorescence signal of the strong transition exhibits bright and dark periods, and the occurrence of a dark period implies that the atom is in $|r\rangle$. Quantum jumps are a good example of how a quantum system far from equilibrium (due to laser driving and spontaneous emission) can have nontrivial dynamics.

The Hamiltonian for a single atom is ($\hbar=1$)
\begin{eqnarray}
H&=&\frac{\Omega_e}{2}(|g\rangle\langle e|+|e\rangle\langle g|)+\frac{\Omega_r}{2}(|g\rangle\langle r|+|r\rangle\langle g|) \nonumber\\
&&-\Delta_e|e\rangle\langle e|-\Delta_r|r\rangle\langle r|,\label{eq:H}
\end{eqnarray}
where $\Delta_e$ and $\Omega_e$ are the laser detuning and driving strength of the strong transition, while $\Delta_r$ and $\Omega_r$ are the corresponding quantities for the weak transition. In the absence of spontaneous emission, Eq.~\eqref{eq:H} would completely describe the system. However, the excited states have lifetimes given by their linewidths, $\gamma_e$ and $\gamma_r$.

In the rest of paper, we make the following assumptions on the parameters. To avoid power-broadening on the strong transition, we choose to work in the low-intensity limit, $\Omega_e\ll\gamma_e$; this choice is clarified in Sec.~\ref{sec:experiment}. For convenience, we set $\Delta_e=0$, although it may be experimentally useful to set $\Delta_e<0$ for continuous laser cooling \cite{metcalf99}. We also set $\gamma_r=0$, since the lifetime of the Rydberg state scales as $n^3$ and hence can be chosen to be arbitrarily long \cite{gallagher94}. It is straightforward to extend the analysis to nonzero $\Delta_e$ and $\gamma_r$.

Well-defined jumps appear in the fluorescence signal when a bright period consists of many photons while a dark period consists of the absence of many photons. For a single atom, this happens when $\Omega_r\ll \Omega_e^2/\gamma_e$ in the case of $\Delta_r=0$ \cite{cohen86}. The transition rate from a dark period to a bright period is \cite{plenio98}
\begin{eqnarray}
\Gamma^{D\rightarrow B}(\Delta_r)&=&\frac{\gamma_e\Omega_e^2\Omega_r^2}{16\Delta_r^4+4\Delta_r^2(\gamma_e^2-2\Omega_e^2)+\Omega_e^4},\label{eq:gammadb}
\end{eqnarray}
and the rate from a bright period to a dark period is
\begin{eqnarray}
\Gamma^{B\rightarrow D}(\Delta_r)&=&\frac{\gamma_e^2+4\Delta_r^2}{\gamma_e^2+2\Omega_e^2}\;\Gamma^{D\rightarrow B}(\Delta_r),\label{eq:gammabd}
\end{eqnarray}
where $B$ and $D$ denote bright and dark periods. The derivation of Eqs.~\eqref{eq:gammadb} and \eqref{eq:gammabd} is reviewed in Appendix A. An important feature of these equations is that both rates are maximum when $\Delta_r=0$, since the strength of the weak transition is maximum there. When $\Delta_r=0$, both rates are approximately $\gamma_e\Omega_r^2/\Omega_e^2$. This depends inversely on $\Omega_e$, because increasing $\Omega_e$ is equivalent to measuring the atomic state more frequently; this inhibits transitions to and from $|r\rangle$, similar to the quantum Zeno effect \cite{itano90}. 

\section{Many-body model}\label{sec:many} Now we consider a one-dimensional chain of $N$ three-level atoms, which are all uniformly excited on the same two transitions. The interatomic distance is assumed to be large enough so that the fluorescence from each atom is resolvable \emph{in situ} on a camera \cite{bakr09}. The atoms are coupled via the dipole-dipole interaction between their Rydberg states. In the absence of a static electric field, the interaction decreases with the third power of distance for short distances and with the sixth power of distance for long distances \cite{saffman10}. We focus on the latter case, since the example numbers given in Sec.~\ref{sec:experiment} are for relatively long distances, although the former case would also be interesting to study. The Hamiltonian is \cite{note1}
\begin{eqnarray}
H&=&\sum_i\Bigg[\frac{\Omega_e}{2}(|g\rangle\langle e|_i+|e\rangle\langle g|_i)+\frac{\Omega_r}{2}(|g\rangle\langle r|_i+|r\rangle\langle g|_i)\nonumber\\
&&-\Delta_r|r\rangle\langle r|_i\Bigg]+\sum_{i<j}\frac{V}{|i-j|^6}|r\rangle\langle r|_i\otimes|r\rangle\langle r|_{j},\label{eq:HN}
\end{eqnarray}
where $V$ is the nearest-neighbor interaction. We have included interactions beyond nearest neighbors in case the long-range interactions are important; it is known that they affect the many-body ground state of Eq.~\eqref{eq:HN} when $\Omega_e=0$ \cite{sela11}.

\begin{figure*}
\centering
\includegraphics[width=7 in]{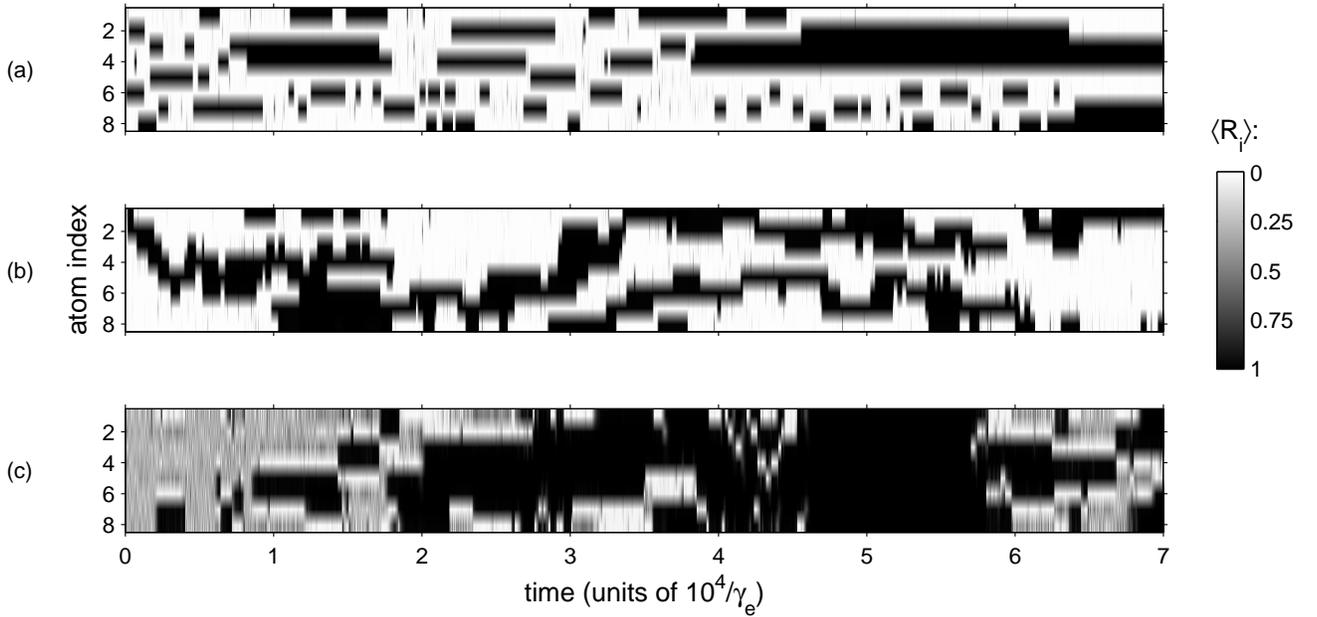}
\caption{\label{fig:n8}Quantum trajectory simulations of a chain of $N=8$ atoms with periodic boundary conditions. The Rydberg population of each atom is plotted vs. time, using color scheme on the right. White color means that the atom is bright and not in the Rydberg state. Black color means that the atom is dark and in the Rydberg state. (a) $\Omega_e=0.2\gamma_e$, $\Omega_r=0.005\gamma_e$, $\Delta_r=0$, $V=0.1\gamma_e$. (b) $\Omega_e=0.2\gamma_e$, $\Omega_r=0.005\gamma_e$, $\Delta_r=V=0.1\gamma_e$. (c) $\Omega_e=\Omega_r=0.1\gamma_e$, $\Delta_r=0$, $V=0.4\gamma_e$.}
\end{figure*}

To demonstrate the rich spatiotemporal dynamics of the many-body system, Fig.~\ref{fig:n8} shows simulations of a chain of $N=8$ atoms, generated using the method of quantum trajectories \cite{dalibard92,molmer93}. Each trajectory simulates a single experimental run. The simulations use periodic boundary conditions and include interactions up to the third neighbor. Figure \ref{fig:n8} plots the time evolution of the Rydberg population of each atom, i.e., the expectation value of $R_i\equiv|r\rangle\langle r|_i$. The atoms undergo quantum jumps, and the Rydberg interaction clearly leads to spatial correlations in the fluorescence.

There are different types of collective dynamics depending on the parameters. In Fig.~\ref{fig:n8}(a)-(b), ${\Omega_r\ll\Omega_e^2/\gamma_e}$, so an atom by itself would exhibit quantum jumps. In Fig.~\ref{fig:n8}(a) ($\Delta_r=0$), a dark period usually does not spread to the neighboring atoms. But once in a while, a dark period does spread to the neighbors, so that there are two or three dark atoms in a row (e.g., $BDDB$). When there are multiple dark atoms in a row, they stay dark for a relatively long time. In Fig.~\ref{fig:n8}(b) ($\Delta_r=V$), once a dark spot is created, it spreads quickly to the neighboring atoms. The dark region expands and contracts in size and appears to diffuse along the chain. Interestingly, when two dark regions are close to each other, they usually do not merge, but appear to ``repel'' each other.
In Fig.~\ref{fig:n8}(c) ($\Omega_r=\Omega_e$, $\Delta_r=0$), the atoms tend to turn dark or bright in groups of two or three, and sometimes all the atoms are dark. The existence of jumps here is surprising because a single atom would not exhibit jumps for these parameters.

\begin{figure*}
\centering
\includegraphics[width=7 in]{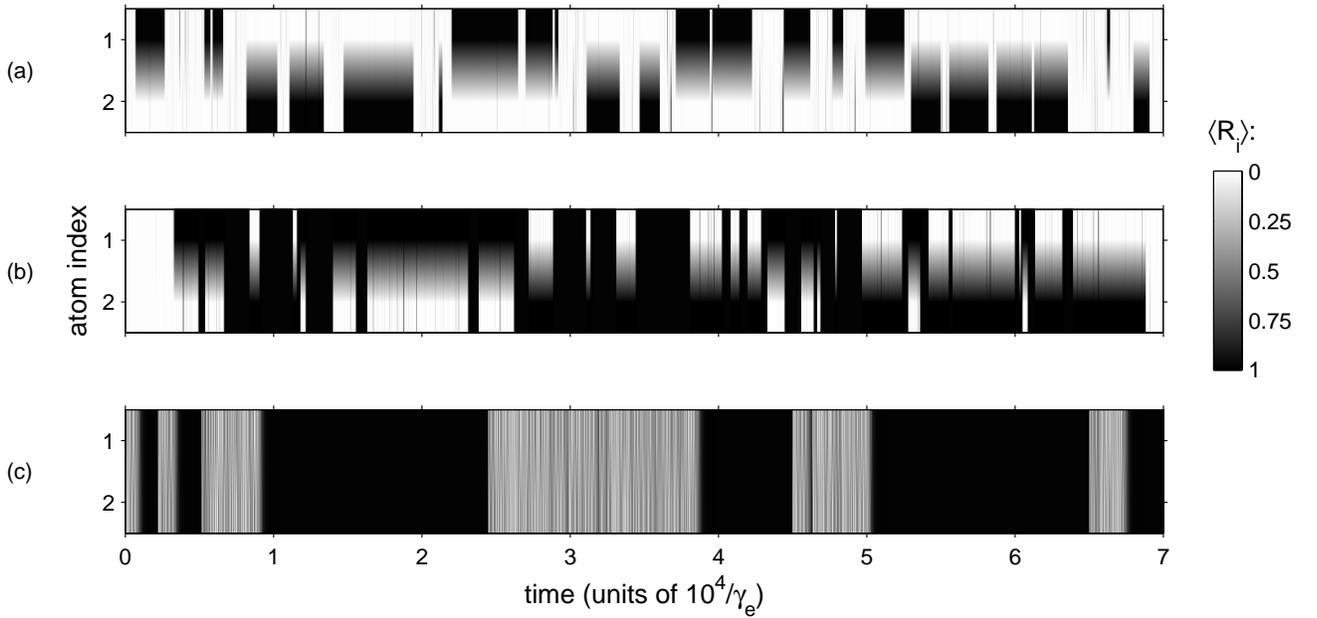}
\caption{\label{fig:n2}Quantum trajectory simulations of $N=2$ atoms. The Rydberg population of each atom is plotted vs. time, using color scheme on the right. Parameters are the same as in Fig.~\ref{fig:n8}: (a) $\Omega_e=0.2\gamma_e$, $\Omega_r=0.005\gamma_e$, $\Delta_r=0$, $V=0.1\gamma_e$. (b) $\Omega_e=0.2\gamma_e$, $\Omega_r=0.005\gamma_e$, $\Delta_r=V=0.1\gamma_e$. (c) $\Omega_e=\Omega_r=0.1\gamma_e$, $\Delta_r=0$, $V=0.4\gamma_e$.}
\end{figure*}

To understand the results for $N=8$, it is instructive to consider the simpler case of $N=2$ atoms. Figure \ref{fig:n2} shows quantum trajectory simulations for $N=2$; note the similarity with Fig.~\ref{fig:n8}. We have analytically solved the $N=2$ case, and the details are in Appendices B and C. In the next two sections, we summarize the $N=2$ results and relate them back to the $N=8$ simulations. There are two general cases: (i) $\Omega_r\ll\Omega_e^2/\gamma_e$ and (ii) $\Omega_r=\Omega_e$, $\Delta_r=0$, distinguished by whether or not a single atom would exhibit jumps.

\section{Case of $\Omega_r\ll\Omega_e^2/\gamma_e$}\label{sec:case1}

\begin{figure}
\centering
\includegraphics[width=3 in]{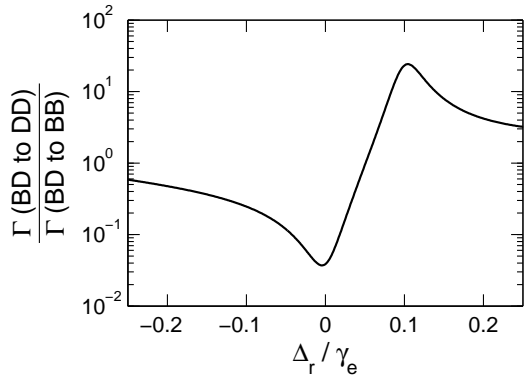}
\caption{\label{fig:n2ratio}Ratio of $\Gamma^{BD\rightarrow DD}$ to $\Gamma^{BD\rightarrow BB}$ for $N=2$ atoms with $\Omega_e=0.2\gamma_e$, $\Omega_r=0.005\gamma_e$, $V=0.1\gamma_e$.}
\end{figure}

For these parameters, an atom by itself would exhibit jumps. Let the two atoms be labelled 1 and 2. If atom 1 is in $|r\rangle$, then according to Eq.~\eqref{eq:HN}, atom 2 effectively sees a laser detuning of $\Delta_r-V$. But if atom 1 is not in $|r\rangle$, then atom 2 sees the original detuning $\Delta_r$. Whether atom 1 is in $|r\rangle$ depends on whether it is in a dark period. This suggests that the jump rates for atom 2 are the same as for a single atom [Eqs.~\eqref{eq:gammadb}-\eqref{eq:gammabd}], except with an effective detuning that depends on whether atom 1 is in a bright or dark period at the moment. In Appendix B, we use a more careful analysis to show that this is indeed correct in the limit of small $\Omega_r$. Thus, the transition rates for two atoms are 
\begin{eqnarray}
\Gamma^{BB\rightarrow BD}(\Delta_r)&=&\Gamma^{BB\rightarrow DB}(\Delta_r)=\Gamma^{B\rightarrow D}(\Delta_r)\label{eq:gbbbd}\\
\Gamma^{BD\rightarrow BB}(\Delta_r)&=&\Gamma^{DB\rightarrow BB}(\Delta_r)=\Gamma^{D\rightarrow B}(\Delta_r)\\
\Gamma^{BD\rightarrow DD}(\Delta_r)&=&\Gamma^{DB\rightarrow DD}(\Delta_r)=\Gamma^{B\rightarrow D}(\Delta_r-V)\\
\Gamma^{DD\rightarrow BD}(\Delta_r)&=&\Gamma^{DD\rightarrow DB}(\Delta_r)=\Gamma^{D\rightarrow B}(\Delta_r-V).\;\;\;\;\label{eq:gddbd}
\end{eqnarray}
An insightful quantity is the ratio $\Gamma^{BD\rightarrow DD}/\Gamma^{BD\rightarrow BB}$, which indicates how often $DD$ periods occur relative to $BB$ periods. As shown in Fig.~\ref{fig:n2ratio}, the ratio is minimum at $\Delta_r=0$ and maximum at $\Delta_r=V$.

The minimum at $\Delta_r=0$ is due to the blockade effect: although the laser is originally on resonance, when atom 1 is in $|r\rangle$, it shifts the Rydberg level of atom 2 off resonance so that atom 2 is prevented from jumping to $|r\rangle$ [Fig.~\ref{fig:levels}(b)]. Thus, the atoms switch between $BB$, $BD$, and $DB$; they are almost never in $DD$. In other words, there is at most one dark atom at a time [Fig.~\ref{fig:n2}(a)]. 

The maximum at $\Delta_r=V$ is due to the opposite effect: the laser is originally off resonance, but when atom 1 happens to jump to $|r\rangle$, it brings the Rydberg level of atom 2 on resonance, encouraging atom 2 to jump to $|r\rangle$ [Fig.~\ref{fig:levels}(c)]. Thus, the atoms switch between $DD$, $BD$, and $DB$; they are almost never in $BB$, except for the initial transient. Since
\begin{eqnarray}\label{eq:ddbd}
\frac{\Gamma^{DD\rightarrow BD}+\Gamma^{DD\rightarrow DB}}{\Gamma^{BD\rightarrow BB}+\Gamma^{BD\rightarrow DD}}&\approx&2,
\end{eqnarray}
a $DD$ period is shorter than a $BD$ or $DB$ period by about a factor of two. When the atoms are in $DD$, there is an equal chance to go to $BD$ or $DB$. Thus, the dark spot appears to do a random walk between the two atoms [Fig.~\ref{fig:n2}(b)]. 

\begin{figure}
\centering
\includegraphics[width=3 in]{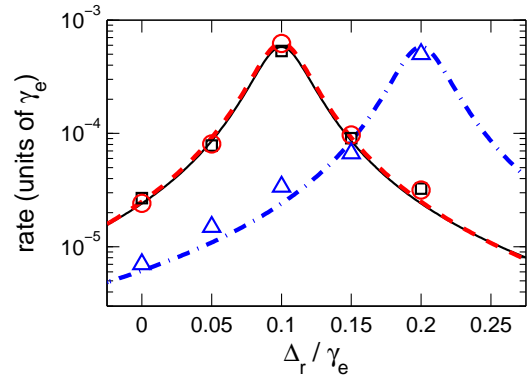}
\caption{\label{fig:n8rates}Dynamics of dark regions in a chain of $N=8$ atoms with $\Omega_e=0.2\gamma_e$, $\Omega_r=0.005\gamma_e$, $V=0.1\gamma_e$. The rates of expansion (black squares), contraction (red circles), and merging (blue triangles) were determined from quantum trajectory simulations. The simulation for each value of $\Delta_r$ was run for a time of $10^6/\gamma_e$, and the rates were calculated by sampling at a rate of $\gamma_e$ and defining an atom to be dark if $\langle R_i\rangle>0.98$. The scatter of data points with low rates is due to statistical uncertainty. Analytical predictions are shown for the rates of expansion (black, solid line), contraction (red, dashed line), and merging (blue, dash-dotted line).}
\end{figure}

The above considerations can be generalized to larger $N$. The transition rates for atom $i$ are given by Eqs.~\eqref{eq:gammadb}-\eqref{eq:gammabd} but with an effective detuning that depends on the number of nearest neighbors that are currently dark: $\Delta_{\text{eff}}=\Delta_r-V\times\mbox{number of dark neighbors}$. This analytical prediction agrees with quantum trajectory simulations of $N=8$ atoms: Fig.~\ref{fig:n8rates} plots the rates of expansion ($\Gamma^{DBB\rightarrow DDB}$), contraction ($\Gamma^{DDB\rightarrow DBB}$), and merging ($\Gamma^{DBD\rightarrow DDD}$) of dark regions. The agreement implies that interactions beyond nearest neighbors in Eq.~\eqref{eq:HN} do not play an important role in the dynamics.

When $\Delta_r=0$, the blockade effect prevents dark periods from spreading [Fig.~\ref{fig:n8}(a)]. But once in a while, a dark period does spread to a neighbor and there are two dark atoms in a row ($BDDB$); when this happens, the dark atoms are effectively off resonance, so they stay dark for a long time. In other words, dark regions expand and contract on a long time scale. Note that the expansion and contraction rates decrease as $V$ increases.

On the other hand, when $\Delta_r=V$, the anti-blockade effect encourages dark periods to spread to the neighbors, causing a dark region to expand [Fig.~\ref{fig:n8}(b)]. But a dark region usually does not expand enough to encompass the entire chain, because an atom at the edge of a dark region can turn bright, causing the dark region to contract. The expansion and contraction processes have similar rates ($\Gamma^{DBB\rightarrow DDB}\approx\Gamma^{DDB\rightarrow DBB}$). As a result, the dark region appears to diffuse randomly along the chain. Also, two dark regions usually do not merge with each other, i.e., $\Gamma^{DBD\rightarrow DDD}$ is relatively small. This is because a bright atom with two dark neighbors is effectively off resonance and is unlikely to turn dark. Hence, the dark regions appear to repel each other.

\section{Case of $\Omega_r=\Omega_e$, $\Delta_r=0$}\label{sec:case2}

\begin{figure}
\centering
\includegraphics[width=3 in]{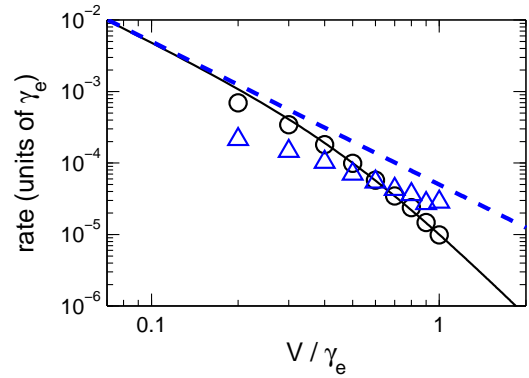}
\caption{\label{fig:n2rates}Jump rates of $N=2$ atoms with $\Omega_r=\Omega_e=0.1\gamma_e$, $\Delta_r=0$. $\Gamma^{DD\rightarrow BB}$: analytical result (black, solid line) and numerical data (black circles). $\Gamma^{BB\rightarrow DD}$: analytical upper bound (blue, dashed line) and numerical data (blue triangles).}
\end{figure}

For these parameters, an atom by itself would not exhibit jumps because of the absence of a weak transition. The existence of jumps for two atoms is solely due to the dipole-dipole interaction, which causes $|gr\rangle\rightarrow|rr\rangle$ and $|rg\rangle\rightarrow|rr\rangle$ to become off-resonant and thus weak transitions [Fig.~\ref{fig:levels}(d)]. Since $|rr\rangle$ is metastable, the system occasionally jumps to and from $|rr\rangle$. When the system is in $|rr\rangle$, the atoms do not fluoresce. When the system is not in $|rr\rangle$, it turns out that the wavefunction rapidly oscillates among the other eigenstates so that both atoms fluoresce from $|e\rangle$. Thus, the system switches between $BB$ and $DD$ [Fig.~\ref{fig:n2}(c)]. In Appendix C, we derive the rates,
\begin{eqnarray}
\Gamma^{DD\rightarrow BB}&=&\frac{\gamma_e\Omega^4}{2V^2(\gamma_e^2+4V^2)}\label{eq:gddbb}\\
\Gamma^{BB\rightarrow DD}&\leq&\frac{\Omega^4}{2\gamma_e V^2},\label{eq:gbbdd}
\end{eqnarray}
where $\Omega\equiv\Omega_r=\Omega_e$. The inequality for $\Gamma^{BB\rightarrow DD}$ is due to incomplete knowledge of the wave function after a photon emission. Equations \eqref{eq:gddbb}-\eqref{eq:gbbdd} agree well with quantum trajectory simulations (Fig.~\ref{fig:n2rates}). Both rates are inversely related to $V$, since the weak transitions become weaker as $V$ increases. The condition for well-defined jumps is roughly $\Omega\ll 2V$.

A larger chain has similar behavior [Fig.~\ref{fig:n8}(c)]. The atoms tend to turn dark or bright simultaneously with their neighbors. However, the dynamics are more complex due to the presence of two neighbors.

\section{Experimental considerations}\label{sec:experiment}

These results can be observed experimentally by using atoms trapped in an optical lattice. For example, one can use ${}^{87}\mbox{Rb}$, which has a strong $5S-5P$ transition with linewidth $\gamma_e/2\pi=6\mbox{ MHz}$ \cite{metcalf99}. Suppose one chooses the $60S$ Rydberg state, which can be reached via a two-photon transition. For a lattice spacing of $7\mbox{ $\mu$m}$, the dipole-dipole interaction decreases with the sixth power of distance \cite{saffman10}, and the nearest-neighbor interaction is $V=0.2\gamma_e$ \cite{reinhard07}. The lifetime of that Rydberg state is $250\mbox{ $\mu$s}$ at 0 K \cite{beterov09}; in other words, $\gamma_r\approx\gamma_e/10^4$. Transitions due to blackbody radiation can be minimized by working at cryogenic temperatures. Also, the $nS$ states have negligible losses from trap-induced photoionization \cite{saffman05,anderson11}. The trapping of Rydberg atoms in optical lattices was recently demonstrated in Refs. \cite{viteau11,anderson11}.

There is an important constraint on the experimental parameters: the interaction $V$ should be much less than the trap depth, or else the repulsive interaction between two Rydberg atoms will push them out of the lattice. Since a trap depth of $10\mbox{ MHz}$ is possible \cite{anderson11}, we require $V\ll\gamma_e$. Then to avoid broadening the strong transition \cite{cohen86} and smearing out the effect of $V$, we choose $\Omega_e\ll\gamma_e$, as stated in Sec.~\ref{sec:single}.

Instead of using the V configuration in Fig.~\ref{fig:levels}(a), one can use a cascade configuration by driving the atom on the $|g\rangle\rightarrow|e\rangle$ and $|e\rangle\rightarrow|r\rangle$ transitions. It is known that quantum jumps occur in this configuration when the upper transition is weak and $|r\rangle$ is metastable \cite{pegg86}. In fact, this is probably the most convenient setup, since experiments often use a two-photon scheme to reach the Rydberg state \cite{wilk10,isenhower10}. To see quantum jumps in a cascade configuration, both transitions should be near resonance instead of far detuned.

\section{Conclusion}

Thus, quantum jumps of Rydberg atoms lead to interesting spatiotemporal dynamics of fluorescence. The next step is to see what happens in larger systems, especially in higher dimensions: what collective behaviors emerge in a large system? It would also be interesting to see what happens when the Rydberg interaction is longer range, i.e., decreasing with the third instead of sixth power of distance; this may lead to significant frustration effects like in equilibrium \cite{sela11}. In addition, one should study what happens when the atoms are free to move instead of being fixed on a lattice; the combination of electronic and motional degrees of freedom will likely result in rich nonequilibrium behavior. Finally, quantum jumps of Rydberg atoms may be a way to experimentally realize the quantum glassiness described in Ref.~\cite{olmos12}. 


We thank H.~H\"{a}ffner and H.~Weimer for useful discussions. This work was supported by NSF Grant No. DMR-1003337.

\appendix
\section{Review of one-atom case}
This appendix reviews the derivation of the jump rates for one atom. We essentially reproduce the derivation in Refs.~\cite{cohen86,porrati89,plenio98}, because we need to refer back to it later, and it is convenient to see it in our notation. In general, we use the ``quantum trajectory'' approach, which is based on the wave function, to account for spontaneous emission.

\begin{figure}
\centering
\includegraphics[width=3 in]{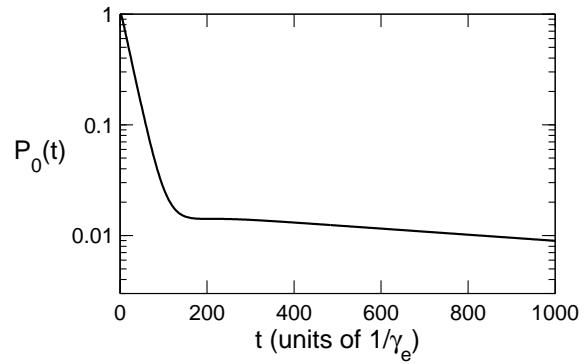}
\caption{\label{fig:p0}Probability that the atom has not emitted by time $t$, given that it emitted at time 0. Parameters are $\Omega_e=0.2\gamma_e$, $\Omega_r=0.2\gamma_e$, $\Delta_e=\Delta_r=\gamma_r=0$.}
\end{figure}

When an atom exhibits quantum jumps, the fluorescence signal has bright periods, in which the photons are closely spaced in time, and dark periods, in which no photons are emitted for a while. The goal is to calculate the transition rate from a bright period to a dark period and vice versa. The important quantity is the time interval between successive emissions \cite{cohen86}. During a bright period, the intervals are short, but a dark period is an exceptionally long interval. Suppose one has the function $P_0(t)$, which is the probability that the atom has not emitted a photon by time $t$, given that it emitted at time 0. $P_0(t)$ decreases monotonically as $t$ increases (Fig.~\ref{fig:p0}). When the parameters are such that there are well-defined quantum jumps, $P_0(t)$ decreases rapidly to a small value for small $t$, but has a long tail for large $t$. This reflects the fact that the time between emissions is usually short (bright period), but once in a while it is very long (dark period). Note that each emission is an independent event, due to the fact that the wave function always returns to $|g\rangle$ after an emission.

We write $P_0(t)=P_{\text{short}}(t)+P_{\text{long}}(t)$ to separate the short and long time-scale parts. The long tail is given by $P_{\text{long}}(t)=p\exp(-\Gamma^{D\rightarrow B}t)$, where $p$ is the probability that a given interval is long enough to be a dark period, and $\Gamma^{D\rightarrow B}$ is the transition rate from a dark period to a bright period. In other words, $1/\Gamma^{D\rightarrow B}$ is the average duration of a dark period.

To calculate $P_0(t)$, we follow the evolution of the wave function $|\psi(t)\rangle$, given that the atom has not emitted a photon yet. This is found by evolving $|\psi(t)\rangle$ with a non-Hermitian Hamiltonian $H_{\text{eff}}=H-i\frac{\gamma_e}{2}|e\rangle\langle e|$. The non-Hermitian term accounts for the population that emits a photon, hence dropping out of consideration \cite{cohen86}. Thus, $P_0(t)=\langle\psi(t)|\psi(t)\rangle$.

In the basis $\{|g\rangle,|e\rangle,|r\rangle\}$, the matrix form of $H_{\text{eff}}$ is
\begin{eqnarray}
H_{\text{eff}}=
\renewcommand{\arraystretch}{1.5}
\left( \begin{array}{ccc}
0 & \frac{\Omega_e}{2} & \frac{\Omega_r}{2} \\
\frac{\Omega_e}{2} & -\frac{i\gamma_e}{2} & 0 \\
\frac{\Omega_r}{2} & 0 & -\Delta_r \end{array} \right).
\end{eqnarray}
As stated in Sec.~\ref{sec:single}, we assume $\Delta_e=\gamma_r=0$. We want to solve the differential equation $i\frac{d}{dt}|\psi(t)\rangle=H_{\text{eff}}|\psi(t)\rangle$
given the initial condition $|\psi(0)\rangle=|g\rangle$. The general solution is $|\psi(t)\rangle=\sum_nc_ne^{-i\lambda_n t}|u_n\rangle$, where $\lambda_n$ and $|u_n\rangle$ are the eigenvalues and eigenvectors of $H_{\text{eff}}$, and $c_n$ are determined from the initial condition $|g\rangle=\sum_nc_n|u_n\rangle$. 

We calculate the eigenvalues and eigenvectors pertubatively in $\Omega_r$, which is assumed to be small. (Note that since $H_{\text{eff}}$ is non-Hermitian, perturbation theory is different from the usual Hermitian case \cite{sternheim72}.) All three eigenvalues have negative imaginary parts, which leads to the nonunitary decay. It turns out that the imaginary part of one of the eigenvalues, which we call $\lambda_3$, is much less negative than the other two. This means that the $|u_1\rangle$ and $|u_2\rangle$ components in $|\psi(t)\rangle$ decay much faster than the $|u_3\rangle$ component. After a long time without a photon emission, $|\psi(t)\rangle$ contains only $|u_3\rangle$. Thus, $\lambda_3$ corresponds to the long tail of $P_0(t)$. 

To second order in $\Omega_r$ \cite{cohen86,porrati89},
\begin{eqnarray}
\lambda_3&=&-\Delta_r+\frac{\Omega_r^2(-2\Delta_r+i\gamma_e)}{8\Delta_r^2-2\Omega_e^2-4i\gamma_e\Delta_r}.
\end{eqnarray}
To first order in $\Omega_r$,
\begin{eqnarray}
|u_3\rangle&=&
\frac{\Omega_r(-2\Delta_r+i\gamma_e)}{4\Delta_r^2-\Omega_e^2-2i\gamma_e\Delta_r}|g\rangle\nonumber\\
&&+\frac{\Omega_e\Omega_r}{4\Delta_r^2-\Omega_e^2-2i\gamma_e\Delta_r}|e\rangle+|r\rangle\label{eq:u3_supp}\\
c_3&=&\frac{\Omega_r(-2\Delta_r+i\gamma_e)}{4\Delta_r^2-\Omega_e^2-2i\gamma_e\Delta_r}.
\end{eqnarray}
Since $|u_3\rangle$ consists mainly of $|r\rangle$, the occurrence of a dark period implies, as expected, that the atom is in $|r\rangle$. (However, note that the atom is not completely in $|r\rangle$. In fact, the dark period ends when the small $|e\rangle$ component in $|u_3\rangle$ decays and emits a photon \cite{plenio98}.)

We can now construct $P_{\text{long}}(t)$:
\begin{eqnarray}
p&=&|c_3|^2\\
&=&\frac{\Omega_r^2(\gamma_e^2+4\Delta_r^2)}{16\Delta_r^4+4\Delta_r^2(\gamma_e^2-2\Omega_e^2)+\Omega_e^4}\\
\Gamma^{D\rightarrow B}&=&-2\;\mbox{Im}\;\lambda_3\\
&=&\frac{\gamma_e\Omega_e^2\Omega_r^2}{16\Delta_r^4+4\Delta_r^2(\gamma_e^2-2\Omega_e^2)+\Omega_e^4}.\label{eq:gdb_supp}
\end{eqnarray}
Then instead of finding $P_{\text{short}}(t)$ explicity, we use a short cut \cite{plenio98}. During a bright period, there is negligible population in $|r\rangle$, so the atom is basically a two-level atom driven by a laser with strength $\Omega_e$. Thus, to lowest order in $\Omega_r$, the emission rate $\Gamma_{\text{short}}$ during a bright period is the same as a two-level atom \cite{metcalf99}:
\begin{eqnarray}
\Gamma_{\text{short}}&=&\frac{\gamma_e\Omega_e^2}{\gamma_e^2+2\Omega_e^2}.\label{eq:gshort}
\end{eqnarray}
However, each emission in a bright period has a small probability $p$ of taking a long time, in which case the bright period ends. Thus, the transition rate from a bright period to a dark period is
\begin{eqnarray}
\Gamma^{B\rightarrow D}&=&p\;\Gamma_{\text{short}}\\
&=&\frac{\gamma_e^2+4\Delta_r^2}{\gamma_e^2+2\Omega_e^2}\;\Gamma^{D\rightarrow B}.\label{eq:gbd_supp}
\end{eqnarray}

The jumps are well-defined when a bright or dark period is much longer than the typical emission time during a bright period: $\Gamma^{B\rightarrow D},\Gamma^{D\rightarrow B}\ll\Gamma_{\text{short}}$. When $\Delta_r=0$ and $\Omega_e\ll\gamma_e$, this condition becomes $\Omega_r\ll\Omega_e^2/\gamma_e$ \cite{cohen86}.

\section{Two atoms, $\Omega_r\ll\Omega_e^2/\gamma_e$}
In this appendix, we derive the jump rates for $N=2$ atoms and $\Omega_r\ll\Omega_e^2/\gamma_e$. For these parameters, a single atom would exhibit quantum jumps. In the case of two interacting atoms, each one still undergoes quantum jumps, but the jump rates of each depend on the current state of the other atom. The goal is to calculate, to lowest order in $\Omega_r$, the transition rates among the possible states: $BB$, $BD$, $DB$, and $DD$.

Suppose for a moment that the interaction strength $V=0$. Then each atom jumps independently, and the jump rates are the same as the single-atom case [Eqs.~\eqref{eq:gdb_supp} and \eqref{eq:gbd_supp}].

Then let $V\neq0$. Due to its form, the Rydberg interaction only affects the state $|rr\rangle$. When the atoms are in $BB$, $BD$, and $DB$, there is negligible population in $|rr\rangle$, so the interaction has negligible effect on the transitions among $BB$, $BD$, and $DB$. So to lowest order in $\Omega_r$, those transition rates are the same as when $V=0$. Thus, we can immediately write down:
\begin{eqnarray}
\Gamma^{BB\rightarrow BD}&=&\Gamma^{BB\rightarrow DB}=\Gamma^{B\rightarrow D}\\
\Gamma^{BD\rightarrow BB}&=&\Gamma^{DB\rightarrow BB}=\Gamma^{D\rightarrow B}.
\end{eqnarray}
The remaining task is to calculate the transition rates that involve $DD$: $\Gamma^{BD\rightarrow DD}$, $\Gamma^{DB\rightarrow DD}$, $\Gamma^{DD\rightarrow BD}$, and $\Gamma^{DD\rightarrow DB}$.

To calculate these rates, we use an approach similar to Appendix A. Suppose the atoms are initially in $BD$, i.e., atom 1 is fluorescing while atom 2 is not. We are interested in the time interval between an emission by atom 1 and a subsequent emission by either atom 1 or 2. Usually the intervals are short since atom 1 is in a bright period. But once in a while, there is a very long interval, which means that atom 1 has become dark and the atoms are in $DD$. If the long interval ends due to an emission by atom 1, the atoms end up in $BD$; if it is due to an emission by atom 2, the atoms end up in $DB$. We want to calculate $P_0(t)$, which is the probability that neither atom has emitted a photon by time $t$, given that atom 1 emitted at time 0 and also given that atom 2 started dark. $P_0(t)$ has a long tail corresponding to time spent in $DD$.

We write $P_0(t)=P_{\text{short}}(t)+P_{\text{long}}(t)$ to separate the short and long time-scale parts. The long tail is given by $P_{\text{long}}(t)=p\exp(-2\Gamma^{DD\rightarrow BD}t)$, where $p$ is the probability that a given interval is long enough to be a $DD$ period. $2\Gamma^{DD\rightarrow BD}$ is the total transition rate out of $DD$ since $\Gamma^{DD\rightarrow BD}=\Gamma^{DD\rightarrow DB}$.

To evolve the wave function in the absence of an emission, we use the non-Hermitian Hamiltonian $H_{\text{eff}}=H-i\frac{\gamma_e}{2}(|e\rangle\langle e|_1+|e\rangle\langle e|_2)$, where $H$ is the two-atom Hamiltonian. We want to solve the differential equation $i\frac{d}{dt}|\psi(t)\rangle=H_{\text{eff}}|\psi(t)\rangle$ in order to find $P_0(t)=\langle\psi(t)|\psi(t)\rangle$.

The question now is what initial condition to use. Since atom 1 is assumed to emit at time 0, it is in $|g\rangle$. Also, as discussed above, during a $BD$ period, there is very little population in $|rr\rangle$, so the interaction has negligible effect on the dynamics. To first order in $\Omega_r$, atom 2's wave function is the same as that of a single atom in a dark period [Eq.~\eqref{eq:u3_supp}]. So the initial condition of the two-atom system is:
\begin{eqnarray}
|\psi(0)\rangle&=&\frac{\Omega_r(-2\Delta_r+i\gamma_e)}{4\Delta_r^2-\Omega_e^2-2i\gamma_e\Delta_r}|gg\rangle\nonumber\\
&&+\frac{\Omega_r\Omega_e}{4\Delta_r^2-\Omega_e^2-2i\gamma_e\Delta_r}|ge\rangle+|gr\rangle.
\end{eqnarray}

The general solution to the differential equation is $|\psi(t)\rangle=\sum_nc_ne^{-i\lambda_n t}|u_n\rangle$, where $\lambda_n$ and $|u_n\rangle$ are the eigenvalues and eigenvectors of $H_{\text{eff}}$, which is a $9\times9$ matrix. $c_n$ are determined from the initial condition $|\psi(0)\rangle=\sum_nc_n|u_n\rangle$. 

We calculate the eigenvalues and eigenvectors perturbatively in $\Omega_r$. All nine eigenvalues have negative imaginary parts, which leads to the nonunitary decay. It turns out that the imaginary part of one of the eigenvalues, which we call $\lambda_9$, is much less negative than the other eight. This means that the other eight components of $|\psi(t)\rangle$ decay much faster than the $|u_9\rangle$ component. After a long time without a photon emission, $|\psi(t)\rangle$ contains only $|u_9\rangle$. Thus, $\lambda_9$ corresponds to the long tail of $P_0(t)$.

To second order in $\Omega_r$,
\begin{eqnarray}
\lambda_9&=&-2\Delta_r+V+\frac{\Omega_r^2(-2\Delta'_r+i\gamma_e)}{4{\Delta'_r}^2-\Omega_e^2-2i\gamma_e\Delta'_r},
\end{eqnarray}
where $\Delta'_r=\Delta_r-V$. To first order in $\Omega_r$,
\begin{eqnarray}
|u_9\rangle&=&
\frac{\Omega_r(-2\Delta'_r+i\gamma_e)}{4{\Delta'_r}^2-\Omega_e^2-2i\gamma_e\Delta'_r}|gr\rangle\nonumber\\
&&+\frac{\Omega_e\Omega_r}{4{\Delta'_r}^2-\Omega_e^2-2i\gamma_e\Delta'_r}|er\rangle\nonumber\\
&&+\frac{\Omega_r(-2\Delta'_r+i\gamma_e)}{4{\Delta'_r}^2-\Omega_e^2-2i\gamma_e\Delta'_r}|rg\rangle\nonumber\\
&&+\frac{\Omega_e\Omega_r}{4{\Delta'_r}^2-\Omega_e^2-2i\gamma_e\Delta'_r}|re\rangle+|rr\rangle\label{eq:u9_supp}\\
c_9&=&\frac{\Omega_r(-2\Delta'_r+i\gamma_e)}{4{\Delta'_r}^2-\Omega_e^2-2i\gamma_e\Delta'_r}.
\end{eqnarray}
Note that $|u_9\rangle$ consists mainly of $|rr\rangle$, since it corresponds to a $DD$ period.

We can now construct $P_{\text{long}}(t)$:
\begin{eqnarray}
p&=&|c_9|^2\\
&=&\frac{\Omega_r^2(\gamma_e^2+4{\Delta'_r}^2)}{16{\Delta'_r}^4+4{\Delta'_r}^2(\gamma_e^2-2\Omega_e^2)+\Omega_e^4}\\
\Gamma^{DD\rightarrow BD}&=&\Gamma^{DD\rightarrow DB}=-\;\mbox{Im}\;\lambda_9\\
&=&\frac{\gamma_e\Omega_e^2\Omega_r^2}{16{\Delta'_r}^4+4{\Delta'_r}^2(\gamma_e^2-2\Omega_e^2)+\Omega_e^4}.\label{eq:gddbd_supp}
\end{eqnarray}
To calculate $\Gamma^{BD\rightarrow DD}$, we use the short cut from Appendix A. Since atom 1 is bright, it has negligible population in $|r\rangle$, so its emission rate $\Gamma_{\text{short}}$ is the same as a two-level atom [Eq.~\eqref{eq:gshort}]. Each emission has probability $p$ of being long enough to be a dark period.
\begin{eqnarray}
\Gamma^{BD\rightarrow DD}&=&\Gamma^{DB\rightarrow DD}=p\;\Gamma_{\text{short}}\\
&=&\frac{\gamma_e^2+4{\Delta'_r}^2}{\gamma_e^2+2\Omega_e^2}\;\Gamma^{DD\rightarrow BD}.\label{eq:gbddd_supp}
\end{eqnarray}
Note the similarity between Eqs.~\eqref{eq:gddbd_supp} and \eqref{eq:gdb_supp} and between Eqs.~\eqref{eq:gbddd_supp} and \eqref{eq:gbd_supp}

\section{Two atoms, $\Omega_r=\Omega_e$, $\Delta_r=0$}
In this appendix, we derive the jump rates for $N=2$ atoms and $\Omega_r=\Omega_e$, $\Delta_r=0$. For these parameters, a single atom would not exhibit quantum jumps. The existence of jumps for two atoms is solely due to the interaction. To calculate the jump rates, we use an approach similar to Appendices A and B, but there are some important differences.

We are interested in the time intervals between photon emissions of either atom. We want to calculate $P_0(t)$, which is the probability that neither atom has emitted a photon by time $t$, given that atom 1 emitted at time 0. (Alternatively, one could let atom 2 emit at time 0.) We write $P_0(t)=P_{\text{short}}(t)+P_{\text{long}}(t)$ to separate the short and long time-scale parts. As in Appendix B, we want to solve the differential equation $i\frac{d}{dt}|\psi(t)\rangle=H_{\text{eff}}|\psi(t)\rangle$ in order to find $P_0(t)=\langle\psi(t)|\psi(t)\rangle$.

Before discussing what initial condition to use, we first calculate the eigenvalues $\lambda_n$ and eigenvectors $|u_n\rangle$ of $H_{\text{eff}}$. We define $\Omega\equiv\Omega_r=\Omega_e$ and do perturbation theory in $\Omega$, which is assumed to be small. As in Appendix B, all nine eigenvalues have negative imaginary parts, which leads to the nonunitary decay. The imaginary part of one of the eigenvalues, which we call $\lambda_9$, is much less negative than the other eight. This means that the other eight components of $|\psi\rangle$ decay much faster than the $|u_9\rangle$ component. Thus, $\lambda_9$ corresponds to the long tail of $P_0(t)$. To fourth order in $\Omega$,
\begin{eqnarray}
\lambda_9&=&V+\frac{\Omega^2}{2V}+\frac{\Omega^4(2V-i\gamma_e)}{4V^2(\gamma_e^2+4V^2)}.
\end{eqnarray}
To first order in $\Omega$,
\begin{eqnarray}
|u_9\rangle&=&\frac{\Omega}{2V}|gr\rangle+\frac{\Omega}{2V}|rg\rangle+|rr\rangle,
\end{eqnarray}
which consists mainly of $|rr\rangle$, reflecting the fact that if both atoms have not emitted for a while, they are in a $DD$ period.

Now it turns out that the real parts of the other eight eigenvalues have very different values, which causes the wave function to oscillate rapidly among the eight eigenvectors. Thus, after atom 1 emits a photon, the short time scale behavior consists of rapid oscillation among the eight eigenvectors, and each atom's $|e\rangle$ population fluctuates a lot. The time scale of the oscillation is faster than the typical photon emission rate, so both atoms are equally likely to emit next. Thus, the atoms can either be in $BB$ or $DD$. When in $BB$, both atoms emit, and the time interval between emissions is relatively short. But once in a while, it takes a very long time for the next photon to be emitted, which means that the atoms are in $DD$. Once the long interval ends, the atoms go back to $BB$.

The rapid oscillation during $BB$ makes it impossible to choose a unique initial condition $|\psi(0)\rangle$, because each time atom 1 emits, atom 2's wave function is different. To account for this ignorance, we let atom 2's wave function be completely arbitrary:
\begin{eqnarray}
|\psi(0)\rangle&=&a_1|gg\rangle+a_2|ge\rangle+a_3|gr\rangle.
\end{eqnarray}
Normalization requires $|a_1|^2+|a_2|^2+|a_3|^2=1$, but $a_1,a_2,a_3$ are otherwise unknown. Despite the incomplete knowledge, we can still obtain a useful bound on $\Gamma^{BB\rightarrow DD}$.

The general solution to the differential equation $i\frac{d}{dt}|\psi(t)\rangle=H_{\text{eff}}|\psi(t)\rangle$ is $|\psi(t)\rangle=\sum_nc_ne^{-i\lambda_n t}|u_n\rangle$, where $c_n$ are determined from the initial condition $|\psi(0)\rangle=\sum_nc_n|u_n\rangle$. To first order in $\Omega$,
\begin{eqnarray}
c_9&=&a_3\frac{\Omega}{2V}.
\end{eqnarray}

Given the above results, we can now construct $P_{\text{long}}(t)=p\exp(-\Gamma^{DD\rightarrow BB}t)$, where $p$ is the probability that a given interval is long enough to be a $DD$ period, and $\Gamma^{DD\rightarrow BB}$ is the transition rate from $DD$ to $BB$:
\begin{eqnarray}
p&=&|c_9|^2\leq\frac{\Omega^2}{4V^2}\\
\Gamma^{DD\rightarrow BB}&=&-2\;\mbox{Im }\lambda_9\\
&=&\frac{\gamma_e\Omega^4}{2V^2(\gamma_e^2+4V^2)}.
\end{eqnarray}
The inequality for $p$ reflects the incomplete knowledge of the initial wave function.

To calculate $\Gamma^{BB\rightarrow DD}$, we have to first calculate $\Gamma_{\text{short}}$, which is the total emission rate of both atoms during a $BB$ period. We approximate $\Gamma_{\text{short}}$ using the emission rate in the absence of the $|g\rangle\rightarrow|r\rangle$ transition, like in Eq.~\eqref{eq:gshort}:
\begin{eqnarray}
\Gamma_{\text{short}}&\approx&\frac{2\gamma_e\Omega^2}{\gamma_e^2+2\Omega^2}.
\end{eqnarray}
However, since the $|g\rangle\rightarrow|r\rangle$ transition is not weak, the above approximation to $\Gamma_{\text{short}}$ is usually an upper bound. Now we can calculate:
\begin{eqnarray}
\Gamma^{BB\rightarrow DD}&=&p\;\Gamma_{\text{short}}\leq\frac{\Omega^4}{2\gamma_e V^2}.
\end{eqnarray}

The jumps are well-defined when a $BB$ period consists of many emissions while a $DD$ period consists of the absence of many emissions: $\Gamma^{BB\rightarrow DD},\Gamma^{DD\rightarrow BB}\ll\Gamma_{\text{short}}$. Roughly speaking, this happens when
\begin{eqnarray}
\Omega&\ll&2V.
\end{eqnarray}

\bibliography{crystal}

\begin{thebibliography}{41}
\bibitem[{not()}]{note1}
\bibinfo{note}{We have taken the rotating-wave approximation and truncated the
  Hilbert space to three levels. The rotating-wave approximation is valid when
  $\Delta_e,\Omega_e\ll\omega_e$ and $\Delta_r,\Omega_r\ll\omega_r$, where
  $\omega_e$ and $\omega_r$ are the transition frequencies of
  $|g\rangle\rightarrow|e\rangle$ and $|g\rangle\rightarrow|r\rangle$. The
  truncation of the Hilbert space is validated by choosing $|e\rangle$,
  $|r\rangle$, and the laser polarizations such that there is negligible
  off-resonant excitation of nearby states. In particular, to avoid an
  accidental jump of the atom to another Rydberg state $|\tilde{r}\rangle$, one
  requires $\Gamma^{B\rightarrow D}(\Delta_{\tilde{r}})\ll\Gamma^{B\rightarrow
  D}(\Delta_r)$.}
\end{thebibliography}

\end{document}